\documentclass[aps,prl,showpacs,twocolumn,superscriptaddress]{revtex4}
\usepackage{bbm}
\usepackage{mathrsfs}
\usepackage{epsfig,psfrag}
\usepackage{amsmath,amsfonts,amssymb}
\usepackage[usenames]{color}

\newcommand{\cZ}{\mathcal Z}
\newcommand{\cD}{\mathcal D}
\newcommand{\cC}{\mathcal C}
\newcommand{\cG}{\mathcal G}
\newcommand{\cE}{\mathcal E}
\newcommand{\br}{\mathbf x}

\newcommand{\bA}{\mathbf A}
\newcommand{\bE}{\mathbf E}
\newcommand{\bB}{\mathbf B}
\newcommand{\bD}{\mathbf D}
\newcommand{\bJ}{\mathbf J}
\newcommand{\bH}{\mathbf H}
\newcommand{\bM}{\mathbf M}

\newcommand{\bQ}{\mathbf Q}
\newcommand{\bP}{\mathbf P}

\newcommand{\bbS}{\mathbb S}
\newcommand{\bbU}{\mathbb U}
\newcommand{\bbT}{\mathbb T}
\newcommand{\bbI}{\mathbb I}
\newcommand{\bbM}{\mathbb M}
\newcommand{\bbN}{\mathbb N}

\newcommand{\bphi}{\mbox{\boldmath$\phi$}}
\newcommand{\bpsi}{\mbox{\boldmath$\Psi$}}

\newcommand{\hr}{\hat{\mathbf x}}
\newcommand{\hz}{\hat{\mathbf z}}

\definecolor{BrickRed}{cmyk}{0,0.89,0.94,0.28}%%%PANTONE 1805
\definecolor{MidnightBlue}{cmyk}{0.98,0.13,0,0.43}%%%PANTONE 302
\definecolor{DarkGreen}{rgb}{0,0.7,0.1}

\def\a{s}
\def\b{s}

\newcommand{\add}[1]{\if\a\b{{\color{magenta} #1}}\else{#1}\fi}
\newcommand{\comm}[1]{\if\a\b{{\color{MidnightBlue}\{\small \sc #1\}}}\else{}\fi}
\newcommand{\del}[1]{\if\a\b{{\color{DarkGreen}[[#1]]}}\else{}\fi}

\begin{document}

\preprint{draft}

\title{Casimir forces between arbitrary compact objects}

\author{T.\ Emig}
\affiliation{Laboratoire de Physique Th\'eorique et Mod\`eles
Statistiques, CNRS UMR 8626, Universit\'e Paris-Sud, 91405 Orsay,
France}

\author{N.\ Graham}
\affiliation{Department of Physics, Middlebury College,
Middlebury, VT  05753} 
\affiliation{
Center for Theoretical Physics, Laboratory for Nuclear Science, and
Department of Physics, Massachusetts Institute of
Technology, Cambridge, MA 02139, USA}

\author{R.\ L.\ Jaffe}
\affiliation{
Center for Theoretical Physics, Laboratory for Nuclear Science, and
Department of Physics, Massachusetts Institute of
Technology, Cambridge, MA 02139, USA}

\author{M.\ Kardar}
\affiliation{
Department of Physics, Massachusetts Institute of
Technology, Cambridge, MA 02139, USA}

\date{\today}

\begin{abstract}
  We develop an exact method for computing the Casimir energy between
  arbitrary compact objects, either dielectrics or perfect
  conductors. The energy is obtained as an interaction between
  multipoles, generated by quantum current fluctuations.  The objects'
  shape and composition enter only through their scattering matrices.
  The result is exact when all multipoles are included, and converges
  rapidly.  A low frequency expansion yields the energy as a series in
  the ratio of the objects' size to their separation.  As an example,
  we obtain this series for two dielectric spheres and the full
  interaction at all separations for perfectly conducting spheres.
\end{abstract}


\maketitle

The electromagnetic (EM) force between neutral bodies is governed by
the coordinated dance of fluctuating charges \cite{Parsegian}.  At the
atomic scale, this attractive interaction appears in the guises of van
der Waals, Keesom, Debye, and London forces.  The collective behavior
of condensed atoms is better formulated in terms of dielectric
properties.  In 1948, Casimir computed the force between two parallel
metallic plates by focusing on the quantum fluctuations of the EM
field between the two plates \cite{Casimir48}.  This was extended by
Lifshitz to dielectric plates, accounting for the fluctuating fields
in the media \cite{Lifshitz56}.  The force between atoms at
asymptotically large distances was computed by Casimir and Polder~
\cite{Casimir+48} and related to the atoms' polarizabilities.  For
compact objects, such as two spheres, Feinberg and Sucher~\cite
{Feinberg+70} generalized this work to include magnetic effects.

In this Letter we obtain the EM Casimir interaction between compact
objects {\em at arbitrary separations} \cite{footnote0}, and determine
explicitly the dependence on shape and material
properties\cite{footnote-1}.  In a qualitative sense, our approach is
similar to a multipole expansion for the fluctuating sources.  The
dependence on shape and material appears through the susceptibility to
current fluctuations, and is related to the scattering of EM waves by
the object. While the scattering matrix is in principle complicated,
there are tools for computing it and it is known for certain
geometries.  As an example, we compute the EM force between two
dielectric spheres at any separation.

Earlier studies of the Casimir force between compact objects include a
multiple reflection formalism\cite{Balian}, which in principle could
be applied to perfect conductors of arbitrary shape.  A formulation of
the Casimir energy of compact objects in terms of their scattering
matrices, for a scalar field coupled to a dielectric background, is
introduced in Ref.~\cite{Kenneth+06}, where it is suggested that it
can also be extended to the EM case.

Many of our results can be derived by either Green's function or path
integral methods.  We shall sketch the latter derivation 
--- due to the letter format only the key steps are outlined, and 
details are left for a
more complete exposition \cite{Emig+07}.  Note first that since the
objects are fixed in time, the action is diagonal in
the frequency $k$. Therefore in all subsequent steps we can treat each
frequency independently, and integrate over $k$ at the end.
The Casimir energy can be associated with modifications of gauge field
fluctuations due to constraints imposed by boundary conditions at 
the material objects.  An alternative and equivalent description,
stressed by Schwinger \cite{Schwinger75}, is to attribute the Casimir
interaction  to fluctuating current and charge densities
$\bJ$, $\varrho$ inside the objects.  In the latter formulation, the
EM gauge and scalar potential $[A(\br,t) ,\Phi(\br,t)]$
are given for each source configuration by the classical solutions,
which in Lorentz gauge read
\begin{equation}
  \label{eq:A-phi-solutions}
  [\bA(\br),\Phi(\br)]=\int d\br' \, G_0(\br,\br') [\bJ(\br'),\varrho(\br')]~,
\end{equation}
with
$G_0(\br,\br')=e^{ik|\br-\br'|}/(4\pi|\br-\br'|)$.   For path integral quantization,
we integrate over all
allowed configurations of the fluctuating currents, weighted by the
appropriate action. The Lagrangian for a collection of currents in
vacuum is  the kinetic energy $\frac{1}{2} \bJ\bA$ minus the
potential energy $\frac{1}{2} \varrho\Phi$. This yields, using
Eq.~\eqref{eq:A-phi-solutions} and the continuity equation
$\nabla \bJ=ik\varrho$, the action $S[\bJ]=\int (dk/4\pi)
(S_k[\bJ]+S^*_k[\bJ])$ for the current densities
$\{\bJ_\alpha\}$ on the objects with
\begin{equation}
  \label{eq:action-currents}
  S_k[\{\bJ_\alpha\}]= \frac{1}{2}\int d \br\,  d \br' \, 
\sum_{\alpha\beta}\bJ_\alpha^*(\br) \, \cG_0(\br,\br') \, \bJ_\beta(\br')\, ,
\end{equation}
where $\cG_0(\br,\br')= G_0(\br,\br') - \frac{1}{k^2} \nabla \otimes
\nabla' G_0(\br,\br')$ is the tensor Green function.  Next we must
constrain the currents to be {\it induced} sources that depend on
shape and material of the objects. Formally this is achieved by
integrating over currents, inserting constraints 
to ensure that the currents in vacuum
simulate the correct induction of microscopic polarization
$\bP_\alpha$ and magnetization $\bM_\alpha$ (from all multipoles)
inside the dielectric objects in response to an incident wave.  

Let us
consider one object.  First, the induced current is
$\bJ_\alpha=-ik\bP_\alpha+\nabla\times \bM_\alpha$, and since
$\bP_\alpha=(\epsilon_\alpha-1)\bE$, $\bM_\alpha=(1-1/\mu_\alpha)\bB$,
it can be expressed in terms of the total fields $\bE$, $\bB$ inside
the object as
\begin{equation}
  \label{eq:J-condition}
  \bJ_\alpha = -ik (\epsilon_\alpha-1) \bE + \nabla \times 
[(1-1/\mu_\alpha) \bB] \, .
\end{equation}
Second, the total field inside the object must consist
of the field generated by $\bJ_\alpha$ and the incident field
$\bE_0(\{\bJ_\alpha, \bbS^\alpha\},\br)$ that has to impinge on the object to
induce $\bJ_\alpha$, so that
\begin{equation}
  \label{eq:total_E}
  \bE(\br)= \bE_0(\{\bJ_\alpha, \bbS^\alpha\},\br) +
 ik \int d \br' \, \cG_0(\br,\br') \, \bJ_\alpha(\br') \, .
\end{equation}
The incident field depends on the current density to be induced and
on the scattering matrix $\bbS^\alpha$ of the object, which connects
the incident wave to the scattered wave. It is fully
specified by the multipole moments of $\bJ_\alpha$ (see below for
details).  Substituting Eq.~\eqref{eq:total_E} and $\bB=(1/ik)
\nabla\times \bE$ into Eq.~\eqref{eq:J-condition} yields
a self-consistency condition that constrains the current
$\bJ_\alpha$. If one writes this condition as
$\cC_\alpha[\bJ_\alpha]=0$ for each object, the functional
integration over the currents constrained this way for all objects
yields the partition function
\begin{equation}
  \label{eq:partition-fct}
  \cZ=\int \prod_\alpha \cD \bJ_\alpha \prod_{\br\in V_\alpha} 
\delta(\cC_\alpha[\bJ_\alpha(\br)])
\exp\left(i S[\{\bJ_\alpha\}]\right) \, .
\end{equation}

It is instructive to look at  two compact objects at a
distance $L$, measured between the (arbitrary) origins ${\cal
  O}_\alpha$ inside the objects.  In this case the action of
Eq.~\eqref{eq:action-currents} is
\begin{eqnarray}
  \label{eq:S-local}
  S_k[\{\bJ_\alpha\}] &=& \frac{1}{2} \sum_{\alpha\neq\beta}
\int d\br_\alpha \, \bJ_\alpha^*(\br_\alpha) \frac{1}{ik} 
\bE_\beta(\br_\alpha-L_\alpha\hz)\\
&+& \frac{1}{2} \sum_\alpha \int \! d\br_\alpha  d\br'_\alpha  \,
\bJ_\alpha^*(\br_\alpha) \, \cG_0(\br_\alpha,\br'_\alpha)\, \bJ_\alpha(\br'_\alpha)
\nonumber \, ,
\end{eqnarray}
where we have substituted the electric field $\bE_\alpha(\br_\alpha)=
ik \int d \br'_\alpha \, \cG_0(\br_\alpha,\br'_\alpha) \,
\bJ_\alpha(\br'_\alpha) $ and the fields are measured now in local
coordinates so that $\br={\cal O}_\alpha+\br_\alpha$, and
$L_\alpha=L$ ($-L$) for $\alpha=1$($2$).  The off-diagonal terms in
Eq.~\eqref{eq:S-local} represent the interaction between the currents
on the two materials.  A natural way to decompose the interaction between 
charges is to use the multipole expansion. For
each body we define magnetic and electric multipoles as
\begin{eqnarray}
  \label{eq:mulitpole-moments}
  Q^\alpha_{\textsc{m},lm}&\!=\!& \frac{k}{\lambda} \!\int\!\! d \br_\alpha \, 
\bJ_\alpha(\br_{\alpha})
\nabla\times[\br_\alpha j_l(kr_\alpha) Y^*_{lm}(\hr_\alpha)]  \\
   Q^\alpha_{\textsc{E},lm}&\!=\!& \frac{1}{\lambda}\!\int\!\! d \br_\alpha \, 
\bJ_\alpha(\br_{\alpha})
\nabla\!\times\!\nabla\times[\br_\alpha j_l(kr_\alpha) Y^*_{lm}(\hr_\alpha)] \nonumber\, ,
\end{eqnarray}
for $l\ge 1$, $|m|\le l$, where $\lambda=\sqrt{l(l+1)}$, $j_l$ are
spherical Bessel functions and $Y_{lm}$ spherical harmonics.  We
change variables from currents to multipoles in the functional
integral and, as the final step in our quantization, integrate over all
multipole fluctuations on the two objects weighted by the effective
action,
\begin{eqnarray}
  \label{eq:S_eff_mp}
  && S^\text{eff}_k[\{Q^\alpha_{lm}\}] =\frac{1}{2} \frac{i}{k} \sum_{lm}\sum_{l'm'} \big\{
Q^{1*}_{lm} \, U^-_{lml'm'} \, Q^2_{l'm'}  \\ 
&+& \!\! Q^{2*}_{lm} \, U^+_{lml'm'} \, Q^1_{l'm'} 
+ \sum_{\alpha=1,2} Q^{\alpha *}_{lm} \, [-T^\alpha]^{-1}_{lml'm'} \, Q^\alpha_{l'm'} 
\big\} \nonumber \, ,
\end{eqnarray}
with $Q^\alpha_{lm}=(Q^\alpha_{\textsc{M},lm},Q^\alpha_{\textsc{E},lm})$.  Let us
discuss the terms appearing in Eq.~\eqref{eq:S_eff_mp} and sketch its
derivation.

{\it Off-diagonal terms} --- We need to know the electric fields in
Eq.~\eqref{eq:S-local} exterior to the source that generates them.
They can be represented in terms of the multipoles as
$\bE_\beta(\br_\beta)=-k \sum_{lm} Q^\beta_{lm}
\bpsi^\text{out}_{lm}(\br_{\beta})$ where
$\bpsi^\text{out}_{lm}(\br_{\beta})$ are \emph{outgoing} vector
solutions of the Helmholtz equation in the coordinates of object
$\beta$ \cite{footnote2}.  We would like to express the currents
$\bJ_\alpha^*$ in Eq.~\eqref{eq:S-local} also in terms of multipoles.
The difficulty in doing so is that the electric field is expressed in
terms of {\it outgoing} partial waves in the coordinates of object
$\beta$, while according to Eq.~\eqref{eq:mulitpole-moments}, the
multipoles involve partial waves $\bpsi^\text{reg}_{lm}(\br_{\alpha})$
that are \emph{regular} at the origin ${\cal O}_{\alpha}$, in the
coordinates of object $\alpha$ \cite{footnote2}.  Going from the
outgoing to the regular vector solutions and changing the coordinate
system involves a translation and change of basis which can be
expressed as $\bpsi^\text{out}_{lm}(\br_\alpha\pm
L\hz)=\sum_{l'm'}U^\pm_{l'm'lm} \bpsi^\text{reg}_{l'm'}(\br_\alpha)$
where the {\it universal} (shape and material independent) matrices
$\bbU^+$ and $\bbU^-$ represent the interaction between the
multipoles. For fixed $(lm)$, $(l'm')$, they are $2\times 2$ matrices
(magnetic and electric multipoles), and functions of $kL$ only. Their
explicit form is known but not provided here to save space
\cite{Wittmann88}; they fall off with $kL$ according to classical
expectations for the EM field. Then the electric field becomes
$\frac{1}{ik}\bE_\beta(\br_\alpha\pm L\hz)= \sum_{lm} \phi^\beta_{lm}
\bpsi^\text{reg}_{lm}(\br_{\alpha})$ with $\phi^\beta_{lm}=i\sum_{lm}
U^\pm_{lml'm'} Q^\beta_{l'm'}$, and the integration in
Eq.~\eqref{eq:S-local} leads, using Eq.~\eqref{eq:mulitpole-moments},
to the off-diagonal terms in Eq.~\eqref{eq:S_eff_mp}.

{\it Diagonal terms} --- The self-action, given by the second term of
Eq.~\eqref{eq:S-local}, is more interesting and more challenging.  It
can be expressed in terms of multipoles if we use the constraint for
the currents, Eqs.~\eqref{eq:J-condition} and \eqref{eq:total_E}. To
do so, we first note that in scattering theory one usually knows the
incident solution and would like to find the outgoing scattered
solution.  They are related by the $S$-matrix.  Here the
situation is slightly different.  We seek to relate a regular solution
$\bE_0(\br_{\alpha})=ik\sum_{lm}\phi_{0,lm}\bpsi^\text{reg}_{lm}(\br_\alpha)$
and the outgoing scattered solution,
$\bE_\alpha(\br_{\alpha})=-k\sum_{lm}Q^{\alpha}_{lm}
\bpsi^\text{out}_{lm}(\br_\alpha)$, generated by the currents in the
material --- a relation determined by the T-matrix,
$\bbT^\alpha\equiv (\bbS^\alpha-\bbI)/2$ --- schematically
$i \bQ^\alpha= \bbT^\alpha\bphi_0$ \cite{S-T-remark,Waterman:1971a}.  We
face the inverse problem of determining $\phi_{0,lm}$ for known
scattering data $Q^\alpha_{lm}$, hence,
\begin{equation}
  \label{eq:incident_amplitudes}
    \phi_{0,lm}= i \sum_{l'm'} \, [T^\alpha]^{-1}_{lml'm'} Q^\alpha_{l'm'} \,
\end{equation}
so that the incident field is given in terms of the S-matrix, as
indicated in Eq.~\eqref{eq:total_E}. Next, we express the self-action
of the currents inside a body (the second term of
Eq.~\eqref{eq:S-local}), as $S^{\alpha}_k[\bJ_{\alpha}] = \frac{1}{2}
\int d \br_\alpha [ \bE \bD^* - \bB \bH^* - ( \bE_0 \bD_0^* - \bB_0
\bH_0^* ) ]$, the change of the field action that results from placing
the body into the {\it fixed} (regular) incident field $\bE_0=\bD_0$,
$\bH_0=\bB_0$, where $\bE$, $\bH$ and $\bD$, $\bB$ are the new total
fields and fluxes in the presence of the body. Using
$\bD=\epsilon_\alpha \bE$, $\bH=\mu_\alpha^{-1}\bB$ inside the body
and Eq.~\eqref{eq:J-condition}, straightforward manipulations lead to
the simple self-action $S^{\alpha}_k[\bJ_{\alpha}]=-\frac{1}{2ik} \int
d \br_\alpha \bJ^*_\alpha \bE_0(\{\bJ_\alpha,\bbS^\alpha\})$.  If we
substitute the regular wave expansion for $\bE_0$ with coefficients of
Eq.~\eqref{eq:incident_amplitudes} and integrate by using
Eq.~\eqref{eq:mulitpole-moments}, we get Eq.~\eqref{eq:S_eff_mp}.

The T-matrix can be obtained for dielectric objects of arbitrary shape
by integrating the standard vector solutions of the Helmholtz equation
in dielectric media over the object's surface \cite{Waterman:1971a}
and both analytical and numerical results are available for many
shapes \cite{T-matrix-refs}.  Hence, for the time being, we shall
assume that the elements of the T-matrix are available.  The
functional integral over multipoles is Gaussian.  The resulting
partition function is an integral over all frequencies of the
determinant of a matrix $\bbM$, with inverse T-matrices along the
diagonal and the matrices $\bbU^\pm$ off the diagonal. For each
$(lm)$, $\bbM$ is a $4\times 4$ matrix (2 polarizations for 2
objects). The generalization to more than two objects is
straightforward.  The result is formally infinite but the infinity can
be trivially removed by dividing by $\cZ_{\infty}$, the partition
function with all objects removed to infinite separations,
corresponding to setting the off-diagonal terms to zero.  Dividing by
$\cZ_{\infty}$ also cancels the functional Jacobian necessary to
transform from an integral over sources to an integral over
multipoles. After a Wick rotation, $k\to i\kappa$, we finally get the
Casimir energy
\begin{equation}
\label{eq:E-final}
  \cE=\frac{\hbar c}{2\pi} \int_0^\infty d\kappa \log \det (\bbI- \bbU^- 
\bbT^2 \bbU^+ \bbT^1) 
\end{equation}
in terms of the matrices introduced in Eq.~\eqref{eq:S_eff_mp}.  
The dependence of the interaction on
distance is completely contained in $\bbU^\pm$, whereas all shape and
material dependence comes from the T-matrices.  With $\bbN\equiv
\bbU^- \bbT^2 \bbU^+ \bbT^1$ it can be written as $\cE=-\frac{\hbar
  c}{2\pi} \int_0^\infty d\kappa \, \text{Tr} \sum_{p=1}^\infty
\frac{1}{p}\bbN^p$ which allows for a simple physical
interpretation. The matrix $\bbN$ scales with distance $L$ as $\sim
\exp(-2L\kappa)$ and describes a wave that travels from one object to
the other and back, involving one scattering at each object. Hence, we
have obtained a multiple-scattering expansion where each elementary
two-scattering process, described by $\bbN$, is further decomposed
into partial waves. This structure allows for a systematic and exact
expansion of the interaction in the inverse distance. At large
distance, the interaction is determined by the small $\kappa$ scaling
of the T-matrix, $T^\alpha_{lml'm'}\sim \kappa^{l+l'+1}$.  This shows
that $2p$ scatterings become important at order $L^{-1-6p}$, and that
partial waves of order $l$ have to be considered at order
$L^{-5-2l}$. Hence, in actual computations, the sum over reflections
can be cut off at finite $p$ and the matrix $\bbN$ can be truncated to
have dimension $2l(2+l)\times 2l(2+l)$ at partial wave order $l$ (see
below). 
We note that Eq.~\eqref{eq:E-final} applies also to spatially varying
but local $\epsilon_\alpha$ and $\mu_\alpha$, since this affects only the
T-matrix. Likewise, it can be extended to any other boundary conditions or
materials by inserting the appropriate T-matrix.

As a specific example, we consider two identical dielectric spheres.
Due to symmetry, the multipoles are decoupled so that the T-matrix is
diagonal, 
\vspace*{-.7cm}
\begin{widetext}
\vspace*{-0.7cm}
\begin{equation}
\label{eq:t-matrix-elem-sphere}
  T^{11}_{lmlm}=(-1)^l \frac{\pi}{2} \frac{\eta I_{l+{1\over 2}}(z)
\left[I_{l+{1\over 2}}(nz)+2nzI'_{l+{1\over 2}}(nz)\right] - n I_{l+{1\over 2}}(nz)
\left[I_{l+{1\over 2}}(z)+2z I'_{l+{1\over 2}}(z)\right]}
{\eta K_{l+{1\over 2}}(z)
\left[I_{l+{1\over 2}}(nz)+2nzI'_{l+{1\over 2}}(nz)\right] - n I_{l+{1\over 2}}(nz)
\left[K_{l+{1\over 2}}(z)+2z K'_{l+{1\over 2}}(z)\right]} \, ,
\end{equation}
\vspace*{-0.2cm} 
\end{widetext}
\vspace*{-1.2cm} 
where the sphere radius is $R$, $z=\kappa R$,
$n=\sqrt{\epsilon(i\kappa)\mu(i\kappa)}$,
$\eta=\sqrt{\epsilon(i\kappa)/\mu(i\kappa)}$, and $I_{l+{1\over 2}}$,
$K_{l+{1\over 2}}$ are Bessel functions.  $T^{22}_{lmlm}$ is obtained
from Eq.~(\ref{eq:t-matrix-elem-sphere}) by interchanging $\epsilon$
and $\mu$. For all partial waves, the {\it leading} low frequency
contribution is determined by the {\it static} electric 
multipole polarizability, $\alpha^\textsc{E}_l =
[(\epsilon-1)/(\epsilon+(l+1)/l)]R^{2l+1}$,
and the corresponding magnetic polarizability, $\alpha^\textsc{M}_l =
[(\mu-1)/(\mu+(l+1)/l)]R^{2l+1}$.  Including the next to leading terms,
the T-matrix has the structure
\begin{equation}
  \label{eq:T-low-kappa}
  T^{11}_{lmlm}=\kappa^{2l}\bigg[\frac{(-1)^{l-1}(l+1)\alpha_l^\textsc{M}}{l (2l+1)!! (2l-1)!!}  \kappa 
+ \gamma^\textsc{M}_{l3}\kappa^{3}+\gamma^\textsc{M}_{l4}\kappa^{4} +\ldots\bigg] \, ,
\nonumber
\end{equation}
and $T^{22}_{lmlm}$ is obtained by $\alpha^\textsc{M}_l \to
\alpha^\textsc{E}_l$,
$\gamma_{ln}^\textsc{M}\to\gamma^\textsc{E}_{ln}$. 
The first terms are
$\gamma^\textsc{M}_{13}=-[4+\mu(\epsilon\mu+\mu-6)]/[5(\mu+2)^2]R^5$,
$\gamma^\textsc{M}_{14}=(4/9)[(\mu-1)/(\mu+2)]^2R^6$, and
$\gamma^\textsc{E}_{13}$, $\gamma^\textsc{E}_{14}$ are obtained again by
the replacement, $\mu\to\epsilon$. Now we can apply our general formula
in Eq.~(\ref{eq:E-final}) to two dielectric spheres with center-to-center
distance $L$. For simplicity, we restrict to two partial waves ($l=2$)
and two scatterings ($p=1$), which yields the exact Casimir energy to
order $L^{-10}$. Matrix operations are performed with {\tt
  Mathematica}, and we find the interaction
\begin{eqnarray}
  \label{eq:2-dielectric-spheres}
&&  \cE=-\frac{\hbar c}{\pi} \bigg\{ \bigg[ \frac{23}{4} \left((\alpha^\textsc{E}_1)^2+
(\alpha^\textsc{M}_1)^2\right) - \frac{7}{2} \alpha^\textsc{E}_1\alpha^\textsc{M}_1 \bigg] \frac{1}{L^7} \nonumber\\
& &+ \frac{9}{16} \big[ \alpha^\textsc{E}_1 \big(59 \alpha^\textsc{E}_2 -11
\alpha^\textsc{M}_2+86 \gamma^\textsc{E}_{13}  -54 \gamma^\textsc{M}_{13} \big)
+ \, \textsc{e} \leftrightarrow \textsc{m} \,  \big] \frac{1}{L^9} \nonumber\\
& &+ \frac{315}{16} \big[  \alpha^\textsc{E}_1 \big( 7 \gamma^\textsc{E}_{14} - 5 \gamma^\textsc{M}_{14} \big)
+ \, \textsc{e} \leftrightarrow \textsc{m} \,
\big] \frac{1}{L^{10}} + \dots \bigg\} \, ,
\end{eqnarray}
where $ \textsc{E} \leftrightarrow \textsc{M}$ indicates 
terms with exchanged superscripts. The
leading term, $\sim L^{-7}$, has precisely the form of the
Casimir-Polder force between two atoms \cite{Casimir+48}, including
magnetic effects \cite{Feinberg+70}. 
The higher order terms are new, and provide the first
systematic result for dielectrics with strong curvature. 
There is no $\sim 1/L^8$ term. 

\begin{figure}[ht]
\includegraphics[width=1\linewidth]{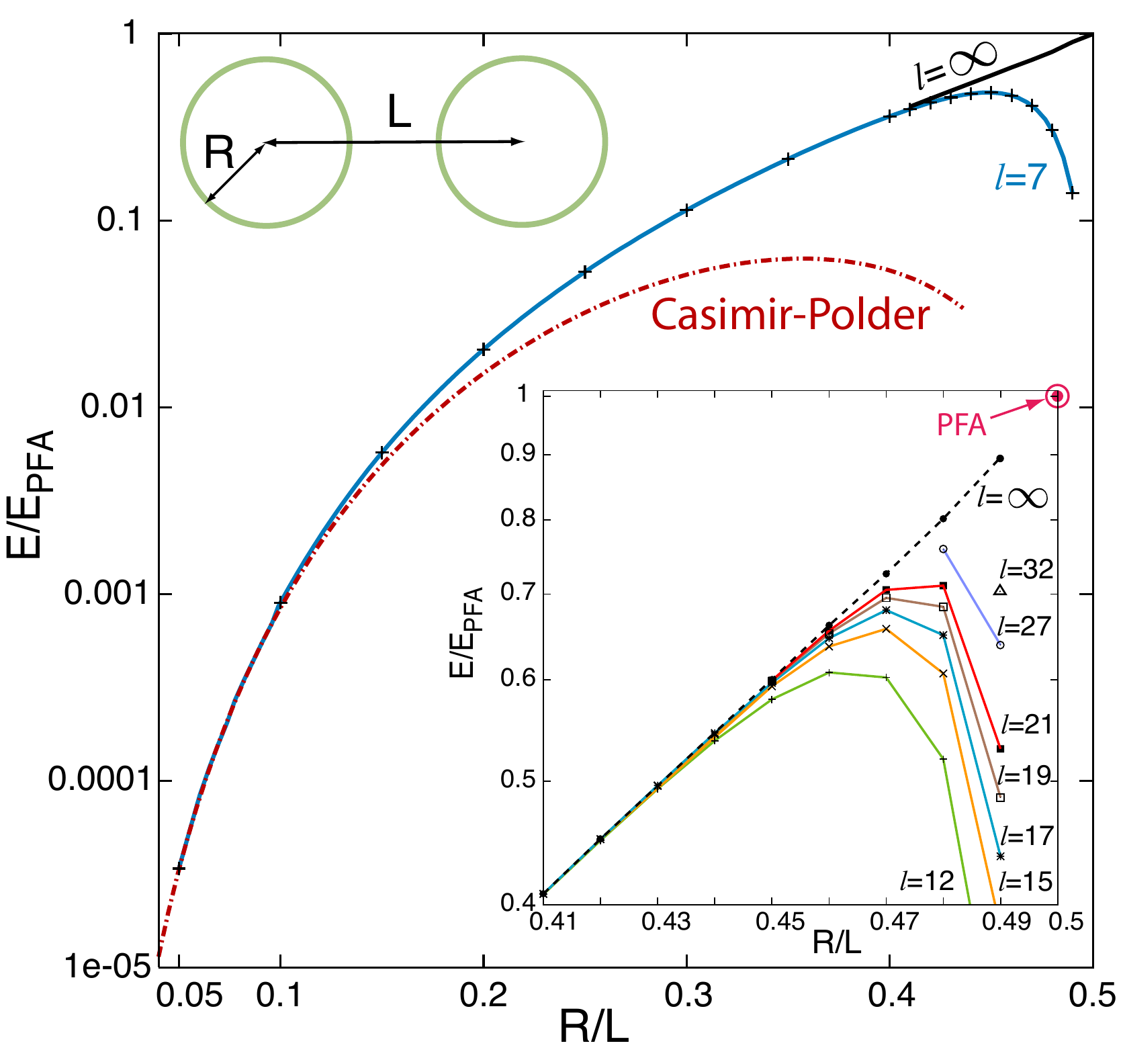}
\vspace*{-.8cm }
\caption{\label{fig:energy}Casimir energy of two metal spheres,
  divided by the PFA estimate $\cE_\text{PFA}=-(\pi^3/1440)\hbar c
  R/(L-2R)^2$, which holds only in the limit $R/L\to 1/2$. The label
  $l$ denotes the multipole order of truncation. The curves $l=\infty$
  are obtained by extrapolation. The Casimir-Polder curve is the
  leading term of Eq.~\eqref{eq:2-metal-spheres}. Inset: Convergence
  at short separations.}
\vspace*{-.6cm }
\end{figure}

The limit of perfect metals 
follows for $\epsilon\to\infty$, $\mu\to 0$. 
Then higher
orders are easily included, yielding an asymptotic series
\begin{equation}
  \label{eq:2-metal-spheres}
  \cE=-\frac{\hbar c}{\pi} \frac{R^6}{L^7} \sum_{n=0}^\infty c_n \left(\frac{R}{L}\right)^n \, ,
\end{equation}
where the first 10 coefficients are $c_0\!\!\!=\!\!\!143/16$,
$c_1\!\!\!=\!\!\!0$, $c_2\!\!\!=\!\!\!7947/160$,
$c_3\!\!\!=\!\!\!2065/32$, $c_4\!\!\!=\!\!\!27705347/100800$,
$c_5=-55251/64$, $c_6=1373212550401/144506880$, $c_7=-7583389/320$,
$c_8=-2516749144274023/44508119040$, $c_9=274953589659739/275251200$.
This series is obtained by expanding in powers of $\mathbb{N}$ and
frequency $\kappa$, and does not converge for any fixed $R/L$.  To
obtain the energy at all separations, one has to compute
Eq.~\eqref{eq:E-final} without these expansions. This is done by
truncating the matrix $\bbN$ at a finite multipole order $l$, and
computing the determinant and the integral numerically. The result is
shown in Fig.~\ref{fig:energy} for perfect metal spheres. Our data
indicate that the energy converges as $e^{-\delta (L/R-2) l}$ to its
exact value at $\l \to \infty$, with $\delta\sim{\cal O}(1)$.  Our
result spans all separations between the Casimir-Polder limit for
$L\gg R$, and the proximity force approximation (PFA) for $R/L \to
1/2$.  At a surface-to-surface distance $d = 4R/3$ ($R/L=0.3$), PFA
overestimates the energy by a factor of 10. Including up to $l=32$ and
extrapolating based on the exponential fit, we can accurately
determine the Casimir energy down to $R/L=0.49$, {\it i.e.\/} $d=0.04
R$.  A similar numerical evaluation can be also applied to dielectrics
\cite{Emig+07}.
%%% Conclusion %%%

We have developed a systematic method for computing the EM Casimir
interaction between compact dielectric objects of arbitrary shapes.
Casimir interactions are completely characterized by the S-matrices of
the individual bodies.  
We have computed the force between spheres for arbitrary
  separations, generalizing previous results that applied only in
  singular limits.  Our method allows for the first time a
description of the Casimir interaction from atomic-scale particles
(Casimir-Polder limit) up to macroscopic objects at short separations
(PFA limit).  For more complicated shapes and multiple objects, it
would be interesting to probe the dependence on the relative
orientations of non-spherical objects and corrections to pair-wise
additivity. Our approach can be applied at finite temperatures and
extended to the computation of correlation functions, energy densities, and the density of
states and may prove also useful to obtain thermal (classical)
fluctuation forces.

This work was supported by the NSF through grants DMR-04-26677 (MK),
PHY-0555338, a Cottrell College Science Award from Research
Corporation (NG), and the U.~S.~Department of Energy (D.O.E.)  under
cooperative research agreement \#DF-FC02-94ER40818 (RLJ).


\begin{thebibliography}{99}
\bibitem{footnote0} The distance is just limited by the standard requirements
on the multipole expansion in classical electrodynamics.
\bibitem{footnote2} The two components 
of  $\bpsi^\text{reg}_{lm}$ are given by $1/k$ times the weights for
  $\textsc{E}$- and $\textsc{M}$-multipoles of
  Eq.~\eqref{eq:mulitpole-moments} with $Y^*_{lm}$ replaced by
  $Y_{lm}$. Similarly, $\bpsi^\text{out}_{lm}$ have the same
  expressions upon substituting Bessel by Hankel functions, $j_l \to
  h_l^{(1)}$.
\end{thebibliography}
\end{document}